\documentclass[conference,a4paper]{IEEEtran}

\usepackage[utf8]{inputenc}
\usepackage[T1]{fontenc}
\usepackage{cite}
\usepackage{amsmath,amssymb,amsfonts}
\usepackage{algorithmic}
\usepackage{graphicx}
\usepackage{textcomp}
\usepackage{xcolor}
\usepackage[hyphens,spaces]{url}
\usepackage{hyperref}
\usepackage{booktabs}
\usepackage{balance}  %
\usepackage{tikz}
\usetikzlibrary{positioning,arrows.meta,calc,fit,backgrounds}
\usepackage{listings}

\def\BibTeX{{\rm B\kern-.05em{\sc i\kern-.025em b}\kern-.08em
    T\kern-.1667em\lower.7ex\hbox{E}\kern-.125emX}}

\title{Agent Flight Recorder: Tamper-Evident Audit Trails
  with On-Chain Anchoring for Long-Horizon Tool-Using Agents}

\author{
    \IEEEauthorblockN{Laurent Bindschaedler\IEEEauthorrefmark{1}, Quentin Botha\IEEEauthorrefmark{2}, Christoph Siebenbrunner\IEEEauthorrefmark{2}}
    \IEEEauthorblockA{\IEEEauthorrefmark{1}Max Planck Institute for Software Systems
    \\bindsch@mpi-sws.org}
    \IEEEauthorblockA{\IEEEauthorrefmark{2}Research Institute for Cryptoeconomics, Vienna University of Economics and Business
    \\\{quentin.botha, christoph.siebenbrunner\}@wu.ac.at}
}

\begin{document}
\pagestyle{empty}

\maketitle

\begin{abstract}
Long-horizon agents execute thousands of actions, resulting in sequential failures rather than isolated errors.
When a coding agent deletes a production database or a prompt injection spreads across agents, the incident raises questions of causality, authority, and non-repudiable third-party verification.

The Agent Flight Recorder captures each agent action as a structured, canonically serialized event binding eight semantic fields from intent through execution to provenance. Hash chaining and Merkle batching provide tamper evidence and compact inclusion proofs. For cross-organizational disputes where no party's infrastructure qualifies as neutral ground, periodic on-chain anchoring of epoch roots lets any verifier with the disclosed payload and Merkle proof check the record independently, without pre-agreeing on a trusted intermediary. The on-chain footprint is minimal: each anchor stores a 32-byte epoch root and a back-pointer, and no event content touches the chain.

We evaluate the system across five cumulative ablation configurations on synthetic agent workloads. The full system adds ${\sim}$48\,$\mu$s median per-event latency and 512~bytes per event. L2 anchoring costs \$2.30 per 100K events at 100-event epochs. The full integrity stack detects edit, delete, reorder, and fork tampering at 100\% with zero false positives. Structured forensic queries achieve 1.0 precision on guardrail and delegation lookups where unstructured text search yields 0.013 and 0.077 respectively.
\end{abstract}


\section{Introduction}
\label{sec:intro}

Long-horizon tool-using agents now operate as autonomous software operators, executing thousands of tool calls over hours or days~\cite{microsoft_governance}. They read documents, call APIs, mutate databases, and delegate tasks to other agents. The dominant failure mode is a wrong sequence of actions, not a wrong answer: a destructive command that ran in production~\cite{replit_incident}, a prompt injection that propagated across agents~\cite{ncsc_prompt_injection}, or an unauthorized privilege escalation~\cite{techradar_second_order}. These failures create operational and legal questions that demand verifiable answers.

Current agent observability stacks (OpenTelemetry spans, vendor dashboards like LangSmith and Langfuse, framework callback logs) help cooperative operators debug failures. They assume mutable internal logs and a single party that both produces and consumes telemetry. They do not provide tamper-evident, externally verifiable records suitable for disputes where the operator itself may be compromised or incentivized to rewrite history.

The Agent Flight Recorder closes this gap with a tamper-evident audit architecture built around an agent-semantic event schema. Each event binds intent, policy evaluation, human approval, execution, effects, context provenance, code provenance, and delegation provenance into a single canonically serialized record. Hash chaining provides per-event ordering and tamper evidence. Merkle batching enables compact inclusion proofs. On-chain anchoring of epoch roots provides a public commitment: any third party holding the disclosed payload and Merkle proof can verify the record independently, with no trust in the operator, no pre-agreed intermediary, and no challenge window.

We evaluate the system across five ablation configurations (Baseline through Full) on synthetic workloads and real SWE-agent~\cite{swe_agent} traces from SWE-bench, deploying the anchoring contract to Base Sepolia and measuring gas from 100~on-chain transactions. We measure per-event latency, storage overhead, tamper detection rates for four attack classes, and forensic query effectiveness on three incident-response scenarios.

The full system adds 48\,$\mu$s median latency per event, detects edit, delete, and reorder attacks at 100\% with zero false positives, and anchors 100K events on an L2 for \$2.30. On-chain anchoring is the only evaluated configuration that enables neutral third-party verification without a pre-agreed intermediary.

This paper makes three contributions: 1) An agent-semantic event schema and tamper-evident audit
    architecture binding intent through execution into canonically
    serialized, hash-chained entries with Merkle batching and on-chain
    anchoring for cross-organizational dispute resolution
    (Section~\ref{sec:design}); 2) A trust-model analysis comparing five anchoring mechanisms
    and identifying when on-chain anchoring provides unique value
    over signed digests, transparency logs, and timestamping
    authorities (Section~\ref{sec:anchoring_analysis}); and 3) An empirical evaluation across five ablation configurations measuring overhead (including validation on real SWE-agent traces), tamper detection, on-chain anchoring cost (measured on Base Sepolia), and forensic query effectiveness (Section~\ref{sec:evaluation}).
The cryptographic primitives are standard. We integrate them into an agent-semantic system that produces an evidentiary settlement architecture, not an observability tool, separating this work from prior tamper-evident logging systems.

\section{Background and Problem Formulation}
\label{sec:background}

\subsection{Agent Actions and Causal Chains}

We define an \emph{agent action} as any discrete step in a long-horizon execution: a tool invocation, an approval gate decision, a delegation request, or an external side effect such as a file write, API call, or database mutation. The unit of harm is a \emph{sequence}, not a single action: what the agent retrieved, what policy the framework checked, what human the system consulted, and in what environment the command executed. An audit system for agents must capture and preserve these ordered causal chains as first-class objects.

\subsection{From Observability to Auditability}
\label{sec:obs_vs_audit}

The gap between observability and auditability is not new. Hash-chain audit logs~\cite{schneier_kelsey}, cloud digest chains~\cite{cloudtrail_integrity}, Merkle-tree transparency logs~\cite{rfc6962}, and blockchain-anchored logging~\cite{putz_blockchain_logging} each address it for different trust models. What is new is that autonomous agents now generate the same category of consequential, disputable actions that previously only human operators and cloud services produced.

\subsection{Threat Model and Assumptions}
\label{sec:threat_model}

We consider an adversary who compromises the agent host at some time
$t_c$. After compromise, the adversary can attempt four operations on the
log: \emph{edit} (modify the payload of an existing entry), \emph{delete}
(truncate or remove entries), \emph{reorder} (swap the sequence of entries),
and \emph{equivocation} (present divergent histories from a common prefix to different verifiers, i.e., a fork attack).

At least one assumption must hold:
\begin{enumerate}
  \item Anchored epoch roots on-chain remain available and immutable (chain
    liveness and finality).
  \item Verifier keys and any external watcher infrastructure are not
    themselves compromised.
  \item An off-host copy of the raw log exists (e.g., replicated storage or
    a remote log sink written before $t_c$).
\end{enumerate}

Following Schneier and Kelsey~\cite{schneier_kelsey}, we draw an explicit
boundary: after full host compromise, the system cannot guarantee the truth
of \emph{newly written} entries. It can only preserve the integrity of
pre-compromise history and detect tampering of entries written before $t_c$.
Anchoring frequency bounds this window but does not eliminate it.

\subsection{Motivating Scenarios}
\label{sec:scenarios}

We ground the design in three concrete scenarios.

\textbf{Scenario 1: Destructive action with disputed postmortem.}
A coding agent deletes a production database~\cite{replit_incident}. The postmortem must determine what tool call ran, what guardrail evaluated it, and whether a human approved.

\textbf{Scenario 2: Prompt injection propagation.}
A malicious instruction in a retrieved document propagates through multiple reasoning steps before triggering a harmful tool call~\cite{ncsc_prompt_injection}. Forensic investigation must trace which retrieved content influenced each action. This requires binding context provenance to each downstream event.

\textbf{Scenario 3: Cross-organizational dispute.}
Two organizations disagree about whether an agent action was authorized. One organization's agent delegated a task to the other's, and the resulting action caused harm~\cite{techradar_second_order}. The audit record must bind ``who asked whom'' across organizational boundaries with a verifiable chain of delegated requests and approvals, and a neutral third party must verify the record without trusting either operator.

\subsection{Positioning Relative to Prior Work}
\label{sec:positioning}

Tamper-evident logging builds on hash chains~\cite{schneier_kelsey}, Merkle-tree proofs~\cite{crosby_wallach}, and blockchain anchoring for generic events~\cite{putz_blockchain_logging}. Recent work applies these ideas to AI agents: an IETF draft~\cite{ietf_agent_audit} defines a hash-chained agent audit format, and other work axiomatizes agentic auditability~\cite{auditable_agentic}. These efforts confirm the problem but leave gaps. The IETF draft has no Merkle batching, no on-chain anchoring, no selective disclosure, and no evaluation. The axiomatic work formalizes properties but does not implement or measure a system.

This paper's system is the first evaluated combination of these layers with an agent-semantic schema, selective disclosure, and on-chain anchoring.

\section{System Design}
\label{sec:design}

\subsection{Design Goals}
\label{sec:goals}

The system targets six goals: \emph{integrity} (no party can modify, delete, or reorder entries without detection); \emph{ordering} (verifiable total event order with concurrency metadata); \emph{external verifiability} (third parties can verify without trusting the operator); \emph{privacy} (payloads encrypted by default, selective per-event disclosure); \emph{low overhead} (sub-millisecond per event, asynchronous anchoring); and \emph{interoperability} (middleware integration, no changes to agent logic). An explicit non-goal is completeness: the recorder guarantees integrity of recorded events, not that all events reach the recorder.

\subsection{Agent-Semantic Event Schema}
\label{sec:schema}

Each event binds eight semantic fields tracing an agent action from intent through execution to effects:

\begin{enumerate}
  \item \textbf{Intent.} The action the model proposed: tool name and
    arguments as structured data. We deliberately exclude raw model
    reasoning and chain-of-thought. Reasoning traces are not stable objects
    (they vary across model providers and versions), raise intellectual
    property and privacy concerns, and model providers may contractually
    prohibit logging them. The recorder captures the \emph{proposed action},
    not the thought process behind it.
  \item \textbf{Policy evaluation.} Which guardrail(s) evaluated the
    proposed action, their verdicts (allow, deny, escalate), and the policy
    version in effect.
  \item \textbf{Human approval.} Whether a human authorized the action,
    the identity of the approver, the approval prompt shown, and the
    approval timestamp. In production deployments, the approval field should carry a cryptographic signature or token binding the approver's identity to the specific action, not merely a metadata string.
  \item \textbf{Execution.} What actually ran: the concrete command, API
    call, or mutation with exact parameters as dispatched to the tool
    runtime.
  \item \textbf{Effects.} What changed: return value, state diff, external
    response, or error. This field captures the observable outcome, not the
    agent's interpretation of it.
  \item \textbf{Context provenance.} The retrieval sources consumed (with
    content hashes), upstream tool outputs referenced, and the environment
    identity (development, staging, production) in which the action
    executed.
  \item \textbf{Code provenance.} The agent binary version hash, tool
    wrapper hashes, and skill or package hashes loaded at
    execution time.
  \item \textbf{Delegation provenance.} For multi-agent workflows: the
    requesting agent's identity, the parent event hash in the requesting
    agent's log, and the scope of delegated authority. This field binds
    ``who asked whom'' across agent boundaries and lets verifiers reconstruct
    cross-agent escalation chains (Scenario~3,
    Section~\ref{sec:scenarios}). Cross-organizational delegation should include a signed delegation token specifying the authority scope, so the verifier can confirm that the delegating party actually authorized the delegated action.
\end{enumerate}

The recorder canonically serializes each event using deterministic Concise Binary Object Representation (CBOR, RFC~8949,
\S4.2). Deterministic CBOR specifies canonical map key ordering and numeric
encoding, eliminating the hash-stability pitfalls of JSON canonicalization
(ambiguous floating-point representation, undefined key ordering, Unicode
normalization variance)~\cite{rfc8785}.
The recorder computes all subsequent hashes over this canonical byte sequence.

\subsection{Cryptographic Construction}
\label{sec:crypto}

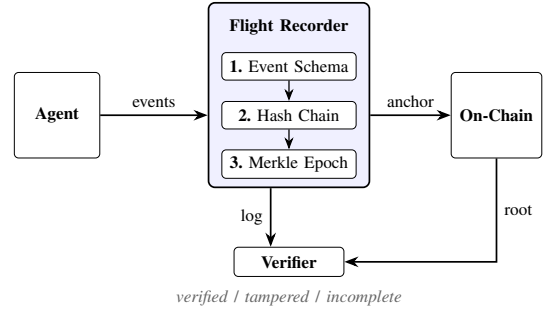
\begin{figure}[t]
\centering
\resizebox{0.8\columnwidth}{!}{\begin{tikzpicture}[
  >=Stealth,
  every node/.style={font=\footnotesize},
  block/.style={draw, rounded corners=2pt, text centered, inner sep=4pt,
                fill=white},
  layer/.style={draw, rounded corners=2pt, minimum width=2.2cm,
                minimum height=0.45cm, text centered, inner sep=3pt,
                fill=white},
  arr/.style={->, thick},
]

\node[block, minimum height=1.4cm, minimum width=1.4cm]
  (agent) at (0, 0) {\textbf{Agent}};

\node[layer] (schema) at (3.8, 0.8)  {\textbf{1.} Event Schema};
\node[layer] (chain)  at (3.8, 0)    {\textbf{2.} Hash Chain};
\node[layer] (merkle) at (3.8, -0.8) {\textbf{3.} Merkle Epoch};

\node[font=\footnotesize\bfseries] (rectitle) at (3.8, 1.45) {Flight Recorder};

\begin{scope}[on background layer]
\node[draw, thick, fill=blue!6, rounded corners=3pt,
      fit=(rectitle)(schema)(chain)(merkle),
      inner xsep=6pt, inner ysep=4pt]
      (recorder) {};
\end{scope}

\node[block, minimum height=1.4cm, minimum width=1.4cm]
  (anchor) at (7.2, 0) {\textbf{On-Chain}};

\node[block, minimum width=1.8cm]
  (verifier) at (3.8, -2.4) {\textbf{Verifier}};

\draw[arr] (agent.east) -- (recorder.west |- agent.east)
  node[above, midway] {events};

\draw[arr] (recorder.east |- anchor.west) -- (anchor.west)
  node[above, midway] {anchor};

\draw[->, semithick] (schema.south) -- (chain.north);
\draw[->, semithick] (chain.south) -- (merkle.north);

\draw[arr] ([xshift=-0.3cm]recorder.south) -- ([xshift=-0.3cm]verifier.north)
  node[left, midway] {log};

\draw[arr] (anchor.south) |- (verifier.east)
  node[right, pos=0.25] {root};

\node[font=\footnotesize, text=black!60, below=0.1cm of verifier]
  {\emph{verified} / \emph{tampered} / \emph{incomplete}};

\end{tikzpicture}}
\caption{Agent Flight Recorder architecture. Each tool call produces a structured event that the recorder serializes, chains, and batches into Merkle epochs. The anchorer periodically commits epoch roots (32~bytes each) to a public chain. The verifier checks event inclusion against the anchored root. Encryption and selective disclosure (Section~\ref{sec:privacy}) are orthogonal to the integrity pipeline and are not shown.}
\label{fig:architecture}
\end{figure}

The architecture, shown in Figure~\ref{fig:architecture}, composes hash chaining for ordering, Merkle batching for compact proofs, and epoch root chaining for gap detection, all finalized via on-chain anchoring.
Removing any layer weakens the system against the corresponding threat class.

We distinguish \emph{event payloads} (the full structured content of an agent action, potentially sensitive) from \emph{event records} (fixed-size metadata: event ID, sequence number, payload hash, predecessor digest, and cleartext metadata such as tool name, timestamp, and recorder identity). All integrity structures (hash chain, Merkle tree, epoch roots) operate on event records. The recorder stores payloads separately and encrypts them when disclosure mode is enabled. This separation keeps the integrity structure free of sensitive content while binding each commitment to a payload via its hash. The cryptographic construction has two local layers and one external layer.

\textbf{Layer~1: Hash chain.}
Each entry $e_i$ includes the hash of its predecessor:
$e_i.\mathit{prev} = H(e_{i-1})$. This establishes a total order over all
recorded events and makes truncation (deletion of a suffix) detectable by
any party holding a later entry. A single-writer path serializes concurrent tool calls. Each event carries a monotonic counter and an optional concurrency-group tag so verifiers can distinguish causal ordering from observation ordering (Section~\ref{sec:implementation}).

\textbf{Layer~2: Merkle batching.}
Every $N$ events form the
leaves of a Merkle tree~\cite{crosby_wallach}. Leaves are SHA-256 event digests, and internal nodes hash ordered pairs. When the leaf count is not a power of two, the implementation duplicates the last leaf. The tree yields $O(\log N)$ inclusion proofs for
individual events and enables efficient batch verification.

\textbf{Epoch root chaining.}
Each epoch $t$ produces a Merkle root $M_t$ over its event digests. The \emph{epoch root} chains this to the previous epoch:
\[
  R_t = H(M_t \;\|\; R_{t-1})
\]
where $R_0 = H(M_0 \;\|\; \mathbf{0}^{32})$. Only $R_t$ goes on-chain, not $M_t$ or any event data, in a digest chain over epochs analogous to CloudTrail's digest files~\cite{cloudtrail_integrity}. Epoch root chaining provides gap detection: if an adversary deletes an entire epoch, the successor's back-pointer will not match any available root. The layers compose: the hash chain prevents tampering within an epoch, the Merkle tree enables compact proofs, and the epoch chain detects gaps between anchoring points.

\subsection{Verification Workflow}
\label{sec:verification}

The verifier checks integrity at two levels, depending on what material it holds.

\emph{Integrity verification (without disclosure).}
Given an event record (which contains the payload hash, not the payload itself), its Merkle proof, and a reference to the on-chain anchor, the verifier can confirm that the event was committed in the anchored epoch without seeing the payload content: (1)~recompute the Merkle path from the event record's hash to the epoch root $R_t$; (2)~confirm $R_t$ matches the on-chain anchor; (3)~verify epoch continuity by checking that each anchored root's back-pointer matches the preceding anchor. This proves the record is authentic and unmodified but reveals nothing about the event's content.

\emph{Content verification (with disclosure).}
When the operator discloses a per-event decryption key (Section~\ref{sec:privacy}), the auditor additionally: (0)~decrypts the ciphertext and verifies that the plaintext hashes to the payload hash inside the committed event record; then proceeds with steps~(1)--(3) above. This proves both authenticity and content integrity.

Both workflows are non-interactive: the verifier needs only the disclosed material and read access to the public chain.

Two requirements sit outside this cryptographic protocol. First, a verifier must resolve \emph{which} contract and recorder identity are authoritative. Each recorder publishes its anchoring contract address and \texttt{msg.sender} identity out-of-band (e.g., in a signed operator manifest). If two anchors claim the same epoch number under that identity, the earlier-included transaction is canonical. A later conflicting anchor from the same identity is evidence of misbehavior, not a valid alternative history. Second, the chain guarantees only that a commitment exists. It does not make the disclosed payload, Merkle proof, or per-event decryption key independently obtainable: those materials require an off-chain data-availability arrangement (a payload store and disclosure policy agreed between the operator and its counterparties or regulator) that is separate from, and not guaranteed by, the anchoring layer itself. On-chain anchoring removes the need to pre-agree on a trusted commitment medium, not the need for a separate availability and disclosure arrangement for the underlying evidence.

\subsection{Selective Disclosure and Privacy}
\label{sec:privacy}

By default, the recorder stores only hashes and structural metadata in the
append-only log. The recorder encrypts and stores full event payloads off-chain.
This ensures that the integrity-critical data structure (hash chain, Merkle
tree, epoch roots) contains no sensitive content, while encrypted payloads
remain available for authorized decryption.

The goal is granular disclosure: an auditor investigating three events out of a thousand should receive decryption keys for only those three, without being able to derive keys for any other event. The system achieves this with a two-level key hierarchy based on HMAC-based Key Derivation Function (HKDF). A single master secret derives a per-epoch key $K_t = \text{HKDF}(\mathit{master},\; \mathit{epoch\_id}_t)$, and each event $i$ within epoch $t$ receives its own encryption key:
\[
  k_{t,i} = \text{HKDF}(K_t,\; \mathit{event\_id}_i)
\]
During an audit, the operator discloses only the per-event keys
$k_{t,i}$ for the specific events under investigation. The auditor
decrypts each corresponding ciphertext and verifies that the plaintext
hashes to the Merkle leaf, confirming both content authenticity and
inclusion in the anchored epoch. Without $K_t$, the auditor cannot derive
keys for other events in the epoch.

This scheme has an explicit limitation: if the auditor obtains the epoch key
$K_t$ itself (e.g., via subpoena or compelled disclosure), they can derive
all per-event keys for that epoch. The construction provides selective
disclosure under \emph{cooperative} audit, not under compelled full
disclosure. A stronger construction using zero-knowledge Merkle inclusion
proofs (proving that an event occurred and matched a policy without
revealing its payload) is a direction for future work.

\section{Anchoring Analysis}
\label{sec:anchoring_analysis}

\subsection{Anchoring Protocol}
\label{sec:anchoring}

At the close of each epoch, the anchorer submits a commitment to a public
blockchain containing the epoch root $R_t$ and the hash of the
previous root $H(R_{t-1})$. The contract records these alongside the caller address (\texttt{msg.sender}, the agent identity), the epoch number (used as the mapping key), and the block timestamp. The on-chain record stores no log
data, only the root and back-pointer.

The architecture follows a pattern analogous to Layer-2 rollups: local execution, batched commitments, and public settlement. The analogy is architectural: the recorder commits to event integrity, not to state-transition correctness.

On-chain anchoring operates as a \emph{non-interactive commitment} model. It differs from optimistic or challenge-based designs, which require a dispute window before finality. The Merkle proof combined with
the on-chain root constitutes a non-interactive proof: any party holding the disclosed payload and proof can
verify independently, at any time, without a dispute window or counterparty interaction once finalized. The system does depend on chain availability for publishing and reading anchors. The anchor
provides a permanent commitment requiring no challenge period and no honest-watcher assumption, unlike optimistic rollups~\cite{optimism_specs} which require a dispute window and at least one honest verifier to submit fraud proofs.
Anchor frequency trades cost for compromise-window width~\cite{schneier_kelsey}: higher frequency narrows the window.

For maximum external verifiability, we target permissionless chains
(Ethereum~L2 or equivalent). Many L2 rollups rely on a single sequencer, a centralization risk for anchor availability: force-inclusion mechanisms let users bypass a censoring sequencer by submitting anchor transactions directly to the rollup's contract on L1, so liveness holds even under sequencer failure. Temporary sequencer unavailability widens the compromise window (the interval between the last anchored state and a potential host compromise) but does not invalidate previously anchored commitments. Anchoring frequency is a tunable parameter, constrained by block time on L1 and by sequencer throughput on L2. Permissioned chains are viable for cost-sensitive deployments where external dispute resolution is not required.

\subsection{Comparison of Anchoring Mechanisms}

Table~\ref{tab:anchoring} compares five anchoring strategies across six
dimensions. We observe that on-chain anchoring is the only
mechanism where pre-agreement between disputing parties is unnecessary,
where any party with the disclosed payload and proof can independently verify the record, and where the
commitment survives operator failure. In all other options, the parties
must have agreed in advance on whose infrastructure to trust: the
operator's signing key, a specific transparency log, or a specific
Timestamping Authority (TSA).

\begin{table*}[t]
\centering
\caption{Comparison of anchoring mechanisms. On-chain anchoring is the only mechanism requiring no pre-agreement and allowing any party to verify given the disclosed payload and proof.}
\label{tab:anchoring}
\scriptsize
\renewcommand{\arraystretch}{0.95}
\begin{tabular*}{\textwidth}{@{\extracolsep{\fill}}l l l l l l@{}}
\toprule
& \textbf{No anchor} & \textbf{Signed digest} & \textbf{Transparency log} & \textbf{TSA} & \textbf{On-chain (ours)} \\
\midrule
Trust assumption            & operator       & signing key    & log operator    & TSA operator     & chain liveness \\
Cross-org disputes          & no             & shared root    & within ecosystem & within PKI      & yes \\
Cost per event              & zero           & negligible     & low             & low              & medium \\
Pre-agreement (commitment medium) & n/a      & signer PKI     & agree on log    & agree on TSA     & none \\
Who can verify              & operator       & pub-key holder & proof holder    & cert holder      & anyone (given proof) \\
Survives operator failure   & no             & if delivered   & if replicated   & if replicated    & commitment \\
\bottomrule
\end{tabular*}
\vspace{2pt}
\raggedright{\scriptsize Independent verification also requires off-chain data availability for the disclosed payload and proof (Section~\ref{sec:verification}).}
\end{table*}

For internal audit, hash chains with write-once read-many (WORM) storage suffice. For bilateral relationships with pre-agreed auditors, a transparency log~\cite{rfc6962} or TSA~\cite{rfc3161} may dominate on cost and simplicity.
Public-chain anchoring becomes justified when the parties have not pre-committed to a neutral intermediary. The next section describes the prototype that implements these layers.

\section{Implementation}
\label{sec:implementation}

The ${\sim}$1,500-line Python prototype has three components.\footnote{Code, evaluation artifacts, and reproduction instructions: \url{https://github.com/mpi-dsg/agent-flight-recorder}.}

\textbf{Local recorder.}
Intercepts tool calls, serializes each into deterministic CBOR, computes the hash-chain pointer and Merkle leaf, and appends to an append-only store. A single-writer path serializes concurrent tool completions. Each event carries a \texttt{logical\_clock} (a monotonic counter, not a vector clock) and an optional \texttt{concurrency\_group} tag so verifiers can distinguish causal from observation ordering. When selective disclosure is enabled, the recorder encrypts each event payload using per-event HKDF-derived keys and stores ciphertexts alongside the integrity structure.

\textbf{Anchorer.}
Batches events into epochs, builds a Merkle tree, and submits the epoch root $R_t$ to an EVM-compatible chain via a ${\sim}$30-line Solidity contract. The contract stores only the 32-byte root and back-pointer on-chain.

\textbf{Verifier.}
Accepts event records (or disclosed payloads), Merkle proofs, and an anchor reference, runs the verification (Section~\ref{sec:verification}), and outputs \emph{verified}, \emph{tampered}, or \emph{incomplete}.

\section{Evaluation}
\label{sec:evaluation}

\subsection{Research Questions}
\label{sec:rqs}

\begin{enumerate}
  \item[\textbf{RQ1.}] What is the latency and storage overhead per tool call?
  \item[\textbf{RQ2.}] What is the on-chain anchoring cost, and how does it
    scale with epoch length?
  \item[\textbf{RQ3.}] Can the system detect all four tamper classes (edit,
    delete, reorder, equivocation/fork) and at which layer?
  \item[\textbf{RQ4.}] How does the recorder affect forensic query
    precision and dispute verifiability for multi-step agent failures?
\end{enumerate}

\subsection{Experimental Setup}
\label{sec:setup}

We use a synthetic workload generator that models a long-horizon agent
executing $N$ tool calls ($N \in \{100, 1\text{K}, 10\text{K}\}$) across
eight tool categories (file read/write, shell command, API call, database
query, retrieval, code execution, approval gate, delegation). We derive category weights from frequency distributions reported in public agent documentation. ${\sim}$15\% of events execute
concurrently (modeled as groups of 2--3 co-issued calls). All runs use seed~42 for reproducibility. The seeded RNG derives all timestamps and random values in the workload, so every event stream is deterministic.

The primary evaluation uses synthetic workloads. We validate RQ1 overhead on five real SWE-agent~\cite{swe_agent} traces from SWE-bench (38~events) and RQ2 anchoring cost on a live Base Sepolia deployment (100~transactions). We run all experiments on a single machine (Apple M-series, 16\,GB RAM, Python~3.11). Each configuration uses the same deterministic event stream for a given seed. We report results across 30 runs with different seeds. The workload generator, tamper injector, and forensic query scripts accompany the artifact.

We evaluate five ablation configurations, each cumulatively adding one layer (Baseline~$\subset$~Schema~$\subset$~Chain~$\subset$~Merkle~$\subset$~Full):
\begin{itemize}
  \item \textbf{Baseline}: plain JSON logging via \texttt{json.dumps},
    no schema, no integrity.
  \item \textbf{Schema}: +~agent-semantic event schema with
    deterministic CBOR serialization.
  \item \textbf{Chain}: +~SHA-256 hash chain.
  \item \textbf{Merkle}: +~Merkle
    batching (100-event epochs).
  \item \textbf{Full}: +~on-chain anchoring.
\end{itemize}
The transition from Baseline to Schema changes both the event structure and the serialization format (JSON to deterministic CBOR), so measured differences reflect the combined effect of schema and encoding, not schema alone. This ablation lets us attribute overhead and detection capability to
each layer independently. Anchoring cost analysis (RQ2) sweeps epoch
sizes of 10, 50, 100, 500, and 1000~events.

We do not include a non-blockchain integrity baseline (e.g., signed digest files or TSA-timestamped logs) in this evaluation. The ablation isolates the value of each layer within our architecture. A head-to-head comparison against alternative anchoring mechanisms would require implementing comparable infrastructure for each. Empirical comparison against deployed alternatives remains future work.

\subsection{Performance Overhead (RQ1)}
\label{sec:perf}

We measure the per-event latency and storage overhead that each integrity layer adds to tool-call recording.

For each ablation configuration and workload size $N$, we record $N \in \{100, 1\text{K}, 10\text{K}\}$ events across 30~independent runs, measure per-event latency via \texttt{perf\_counter\_ns}, and compute storage from the serialized log size. Table~\ref{tab:overhead} reports results at $N = 10\text{K}$.

\begin{table}[htbp]
\centering
\caption{Per-event overhead across ablation configurations}
\label{tab:overhead}
\scriptsize
\renewcommand{\arraystretch}{0.95}
\begin{tabular}{l r r r r}
\toprule
Configuration & Median ($\mu$s) & P95 ($\mu$s) & P99 ($\mu$s) & Bytes/event \\
\midrule
Baseline & 20.3 & 26.2 & 30.7 & 272 \\
Schema & 6.0 & 8.2 & 10.0 & 476 \\
Chain & 48.2 & 57.2 & 76.8 & 477 \\
Merkle & 48.8 & 60.8 & 4067.6 & 512 \\
Full & 47.6 & 60.2 & 30020.9 & 512 \\
\bottomrule
\end{tabular}
\end{table}

Schema (CBOR) is faster than Baseline (JSON) at the median (6.0 vs.\ 20.3\,$\mu$s) because deterministic CBOR encoding of structured dataclass fields avoids the overhead of Python's \texttt{json.dumps} on nested dicts. The trade-off is 75\% more bytes per event (476 vs.\ 272). Adding the hash chain (Chain) raises the median to 48.2\,$\mu$s, dominated by SHA-256 hashing and chain-pointer bookkeeping. Merkle batching adds negligible median cost (48.8\,$\mu$s) but introduces a P99 spike to ${\sim}$4.1\,ms because every 100th event triggers epoch sealing, which builds the Merkle tree and chains epoch roots. The Full configuration adds ECDSA transaction signing at each epoch seal. The median (47.6\,$\mu$s) is within noise of Merkle because signing occurs only at epoch boundaries (1\% of events), but the P99 rises to ${\sim}$30\,ms, reflecting the cost of building and signing an anchor transaction. In a production deployment, the signed transaction would be submitted asynchronously. The P99 here measures the local preparation cost, not the network round-trip. Storage scales from 272~bytes/event (Baseline) to 512~bytes/event (Full), an 88\% increase attributable to the structured schema, hash pointers, and epoch metadata.

\textbf{Takeaway.} The full integrity stack adds ${\sim}$48\,$\mu$s median per-event latency and 512~bytes per event, well within the latency budget of any tool call that involves I/O. Transaction signing for on-chain anchoring adds ${\sim}$26\,ms at epoch boundaries (P99), a cost that is amortized over the epoch and asynchronous in production.

\textbf{Validation on real agent traces.} We replay five SWE-agent~\cite{swe_agent} trajectories from SWE-bench tasks (38~events total; 47\%~shell commands, 34\%~file edits, 18\%~file reads) through the same five configurations. The relative pattern holds: Full adds ${\sim}$14\,$\mu$s median over Baseline (2.4\,$\mu$s), with signing overhead at P95. Absolute latencies are lower because real events have fewer populated fields than the synthetic workload's eight-category distribution. The tool-type distribution differs from the synthetic assumption (no delegation or approval events in coding tasks).

\subsection{Anchoring Economics (RQ2)}
\label{sec:cost}

We measure the on-chain anchoring cost and its tradeoff against the compromise window.

We deploy the anchoring contract to Base Sepolia and measure gas consumption over 100~anchor transactions. The mean gas per \texttt{anchor()} call is 91,800 (71,680 for the first cold-storage write, 91,976 for subsequent warm writes). We sweep epoch sizes of 10, 50, 100, 500, and 1000 events for a workload of 100K events. We project costs using the measured gas at 30~Gwei on Ethereum~L1 and 0.01~Gwei on an L2, with ETH at \$2,500. The compromise window equals $\text{epoch\_size} \times \text{mean\_event\_interval}$, modeled at one event per second. Table~\ref{tab:anchoring_cost} reports the results.

\begin{table}[htbp]
\centering
\caption{Anchoring cost vs.~compromise window (assuming 1~event/s; bursty workloads shift the wall-clock window)}
\label{tab:anchoring_cost}
\scriptsize
\renewcommand{\arraystretch}{0.95}
\begin{tabular}{r r r r r}
\toprule
Epoch size & Window (s) & Anchors & L1 cost (\$) & L2 cost (\$) \\
\midrule
10 & 10 & 10000 & 68850.0000 & 22.950000 \\
50 & 50 & 2000 & 13770.0000 & 4.590000 \\
100 & 100 & 1000 & 6885.0000 & 2.295000 \\
500 & 500 & 200 & 1377.0000 & 0.459000 \\
1000 & 1000 & 100 & 688.5000 & 0.229500 \\
\bottomrule
\end{tabular}
\end{table}

L2 anchoring reduces costs by $3{,}000{\times}$ relative to L1: at epoch size~100, anchoring 100K events costs \$6,885 on L1 but only \$2.30 on an L2 rollup. The cost scales linearly with the number of anchors, which is inversely proportional to epoch size. At the smallest epoch size (10), the compromise window is just 10~seconds but L1 cost rises to \$68,850. At epoch size~1000, L1 cost drops to \$689, but the compromise window widens to ${\sim}$17~minutes. In practice, agent workloads are bursty, so the wall-clock compromise window depends on workload patterns, not just epoch size. The L2 cost curve makes sub-minute compromise windows feasible: even at epoch size~10, L2 cost is \$22.95 for 100K events.

\textbf{Summary.} L2 anchoring costs \$2.30 per 100K events at 100-event epochs (measured gas, projected fees), making sub-minute compromise windows practical.

\subsection{Tamper Detection (RQ3)}
\label{sec:tamper}

We test whether the system detects all four attack classes (edit, delete, reorder, equivocation/fork) and at which ablation layer detection begins.

We inject each attack at ten positions across a 10,000-event log: eight interior positions spread evenly through the log plus the two terminal positions (indices $N{-}2$ and $N{-}1$), which are the most security-critical because they have the fewest successors to detect chain breaks. We run the verifier and record whether tampering is detected. Table~\ref{tab:detection} reports detection rates. False positive rate is 0\% across configurations and attack types.

\begin{table}[htbp]
\centering
\caption{Tamper detection rates by attack type and configuration}
\label{tab:detection}
\scriptsize
\renewcommand{\arraystretch}{0.95}
\begin{tabular}{l r r r r r}
\toprule
Attack & Baseline & Schema & Chain & Merkle & Full \\
\midrule
Edit & 0\% & 0\% & 90\% & 100\% & 100\% \\
Delete & 0\% & 0\% & 90\% & 100\% & 100\% \\
Reorder & 0\% & 0\% & 100\% & 100\% & 100\% \\
Equiv./fork & 0\% & 0\% & 0\% & 0\% & 100\% \\
\bottomrule
\end{tabular}
\end{table}

Baseline and Schema detect nothing: without hash chaining, there is no integrity mechanism to break. Chain detects edit and delete at 90\% (9/10 positions) and reorder at 100\%. The one missed position is the log's final event ($N{-}1$): editing or deleting it leaves no successor whose \texttt{prev\_hash} would expose the break. The penultimate event ($N{-}2$) \emph{is} detected because event $N{-}1$ still checks its predecessor's hash. Merkle closes this gap (100\% for all single-log attacks) because the epoch's committed leaf hashes catch any payload modification: the tampered event's hash no longer matches the leaf committed when the epoch was sealed. Chain and Merkle are identical for reorder detection since swapping adjacent events always breaks at least one interior chain link.

Equivocation (fork) warrants separate treatment. A forked chain is internally valid on each branch, so single-log verification cannot detect it. Detection requires a \emph{dispute-verification protocol}: we record the original branch with anchoring, then record a forked branch through a separate recorder and check the forked epochs against the original's anchored roots. In the Full configuration, every forked epoch produces a different root that mismatches the anchor, yielding 100\% detection across all ten fork positions. Without anchoring, both branches are self-consistent and equivocation is undetectable.

\textbf{Takeaway.}
The system achieves 100\% detection across all attack classes. The Merkle layer secures terminal events, and anchoring enables equivocation detection.

\subsection{Forensic Queryability (RQ4)}
\label{sec:drills}

We test how the recorder affects forensic query precision and dispute verifiability for multi-step agent failures.

We plant failures into three synthetic scenarios (Section~\ref{sec:scenarios}) and run forensic queries across all five configurations. Scenario~A (prompt injection, five queries: poisoned-document identification, subsequent calls, guardrail evaluation, delegation tracing, causal chain) and Scenario~B (production mis-targeting) test the agent-semantic schema. Scenario~C (cross-organizational dispute) tests on-chain anchoring. Precision differs sharply across configurations.

\emph{Scenario~A (prompt injection).}
Baseline's text grep for guardrail evaluations (Q3) returns 79 matches for one true positive (precision~1.3\%), because the regex \texttt{policy\_eval|guardrail} matches unrelated events. Schema through Full use exact field lookups (\texttt{policy\_eval.rule == "destructive\_command\_check"}) and achieve 1.0 precision and recall on all five queries. The delegation query (Q4) exposes a second structural advantage. ``Was the destructive action delegated?'' requires following the \texttt{delegation.parent\_event} pointer back to the injection chain, a relational lookup that structured search handles directly (precision~1.0). Grep can only approximate this by matching the keyword \texttt{delegation}, hitting all 13~delegation events (precision~0.077). Baseline cannot reconstruct the full causal chain (Q5: 0.0 precision/recall) because unstructured logs lack the hash-chain pointers needed to link events into a sequence.

\emph{Scenario~B (production mis-targeting).}
Baseline's grep for ``staging'' returns two events (precision~0.5) while structured search on \texttt{context\_provenance.environment} returns exactly the one configuration-read event.

\emph{Scenario~C (cross-organizational dispute).}
All configurations can resolve the dispute from the logs (Q1), but only the Full configuration with on-chain anchoring enables neutral third-party verification~(Q2): without a public commitment, the verifier must trust one party's key or log server.

\textbf{Finding.} Structured fields eliminate false positives in forensic triage (1.0 vs.\ 0.013 precision on guardrail queries, 1.0 vs.\ 0.077 on delegation queries). On-chain anchoring is the only configuration enabling neutral third-party dispute resolution.

\section{Related Work}
\label{sec:related}

\subsection{Secure Logging Foundations}

Hash-chain techniques make log entries generated prior to compromise resistant to undetectable modification~\cite{schneier_kelsey}, a property later formalized as forward integrity~\cite{bellare_yee_forward_integrity}. Efficient Merkle-tree structures give tamper-evident logging with logarithmic proof sizes~\cite{crosby_wallach}, and recent work co-designs a high-performance tamper-evident logging system with eBPF~\cite{zhao_nitro}. We inherit the Schneier--Kelsey threat boundary.

AWS CloudTrail~\cite{cloudtrail_integrity} provides digest-based integrity for cloud logs but trusts AWS as the operator. That assumption fails when the operator is a disputing party.

\subsection{Transparency Logs and Public Commitment}

Certificate Transparency~\cite{rfc6962} established Merkle-tree inclusion proofs for append-only auditability. Sigstore~\cite{newman_sigstore}, Trillian~\cite{trillian}, RFC~3161 timestamping~\cite{rfc3161}, and keyless-signature infrastructure~\cite{buldas_ksi} generalize it. Each improves auditability but relies on a designated log ecosystem or authority. Section~\ref{sec:anchoring_analysis} compares these systematically.

\subsection{Blockchain-Based Auditable Logging}

A two-layer pattern of local hash chaining plus periodic on-chain commitment, which we extend (Section~\ref{sec:positioning}), appears in blockchain-assisted audit logging~\cite{putz_blockchain_logging}. Related systems store log hashes on-chain for integrity~\cite{pourmajidi_logchain} or use Hyperledger Fabric for runtime governance of LangChain-based agents~\cite{jan_blockchain_agent}. All three target generic events or coarse-grained governance rather than per-action agent-semantic audit.

\subsection{Agent Observability and Audit}

Mainstream agent tracing platforms (OpenTelemetry spans, LangSmith, Braintrust, Langfuse) capture prompts, tool spans, and execution flows for debugging and monitoring. These platforms assume a cooperative operator and do not treat the log as an externally committed evidence object.

Recent projects target tamper-evident agent logs specifically. The AI Action Ledger~\cite{ai_action_ledger} uses hash chaining with privacy-by-default, while the AIR project~\cite{air_project} motivates the problem as ``logs are scattered, postmortems are hard.'' ProofTrail~\cite{prooftrail} provides Merkle proofs for tool-call receipts, and a lifecycle framework for LLM audit trails covers governance, training, and deployment~\cite{ojewale_audit_trails}.

An IETF Internet-Draft~\cite{ietf_agent_audit} defines a JSON-based agent audit format with SHA-256 hash chaining and optional ECDSA signatures, while work on axiomatic auditability~\cite{auditable_agentic} formalizes integrity, coverage, and verifiability properties for agentic systems. A recent SoK systematizes security and privacy concerns for AI agents on blockchains~\cite{bcca2025_sok}. Prior work does not jointly provide (i)~an agent-semantic schema binding intent through execution, (ii)~external anchoring for cross-organizational disputes, (iii)~selective per-event disclosure, and (iv)~empirical ablation on long-horizon agent failures.

On the regulatory side, Microsoft Purview~\cite{microsoft_purview} provides audit logging for Copilot interactions. Article~12 of the EU AI Act mandates automatic event recording for high-risk AI systems but does not prescribe blockchain or cryptographic tamper-evidence~\cite{eu_ai_act}.

\section{Discussion}
\label{sec:discussion}

\subsection{Limitations}

Six limitations bound the system's guarantees.
\emph{(1)~Completeness.} If an action bypasses the instrumented gateway, the recorder never sees it. Within-epoch suppression is undetectable from the log alone. The epoch chain detects missing epochs (broken back-pointers), but not missing events within an epoch. This is a fundamental boundary of software-based audit. Binding the tool-invocation gateway to a remotely attested trusted execution environment (TEE) would change this guarantee: instead of trusting that the operator routed every action through the recorder, a verifier could check a remote-attestation quote proving the gateway ran an unmodified, known-good build. This makes bypass structurally harder rather than merely policy-prohibited. This does not make completeness absolute (a compromised attestation key or a side channel outside the attested boundary remains a residual risk). However, it shifts the assumption from ``the operator did not disable instrumentation'' to ``the attested code measurement matches a published build,'' which a third party can check independently. We treat this as future work (Section~\ref{sec:future}) rather than a property of the current system.
\emph{(2)~Post-compromise truth.} Per Schneier and Kelsey~\cite{schneier_kelsey}, entries written after full host compromise cannot be trusted (Section~\ref{sec:threat_model}).
\emph{(3)~Payload recovery.} The verifier needs the event payload and Merkle proof, not just the on-chain root. If the payload store is lost with no off-host copy, the commitment proves a log existed but content is unrecoverable.
\emph{(4)~Side channels.} The recorder captures tool-call semantics, not model internals or out-of-band agent communication.
\emph{(5)~Anchoring availability.} When the chain is unavailable, actions proceed unanchored and the compromise window expands until connectivity resumes.
\emph{(6)~Strategic anchoring delay.} An operator may delay anchoring. External monitoring of anchor regularity mitigates this but requires a watcher.

\subsection{Privacy and Compliance}

Cryptographic erasure (Section~\ref{sec:privacy}) requires destroying or re-wrapping the relevant epoch-key material, not merely deleting derived per-event keys, since the HKDF derivation is deterministic from the epoch key. Whether such erasure satisfies GDPR depends on jurisdiction, metadata linkability, and retention policy. We present it as an operational mechanism rather than a legal sufficiency claim. The EU AI Act (Article~12) mandates automatic event recording for high-risk AI systems~\cite{eu_ai_act}. The flight recorder's schema supports this, and selective disclosure balances accountability against confidentiality.

\subsection{Threats to Validity}

\emph{Internal.} The primary workloads are synthetic. Tool-type weights draw from public documentation. We validate RQ1 on five real SWE-agent traces from SWE-bench, confirming the overhead pattern holds. RQ1 uses 30 runs, while RQ3 and RQ4 use single deterministic runs. Reported detection rates apply under the stated evidence model, and suffix truncation after compromise requires off-host copies.
\emph{External.} Gas prices fluctuate. Our dollar figures are illustrative under stated assumptions.
\emph{Construct.} The drills use experimenter-defined queries on simulated failures rather than independent incident responders on real incidents, and the baseline is plain JSON logging rather than a mature audit system. The ablation isolates component-level effects rather than claiming superiority over existing platforms. Fork detection is not a single-log property: our tamper table covers single-log verification only. Forensic query improvements partly reflect the query backend and should not be generalized beyond the tested configuration.
\emph{Operational.} Canonicalization edge cases, key management failures, adversarial inputs, anchor delay under chain congestion, and reorg/finality edge cases could undermine guarantees but do not change the cryptographic properties measured.

\subsection{Future Directions}
\label{sec:future}

Five directions merit investigation:
(1)~ZK proofs over Merkle inclusion for privacy-preserving compliance under compelled audit;
(2)~TEE-backed completeness via remote attestation of the tool gateway;
(3)~gas-adaptive anchoring that widens epochs when fees spike;
(4)~on-chain bond mechanisms where recorders stake collateral claimable via fraud proof, ensuring someone \emph{does} verify what anyone \emph{can}; and
(5)~a cross-vendor Agent Transparency Log standard analogous to Certificate Transparency.

\section{Conclusion}
\label{sec:conclusion}

Long-horizon agent failures are sequence failures, and resolving them under dispute requires tamper-evident, neutral evidence that no single party controls. The Agent Flight Recorder provides this by combining an agent-semantic schema with hash chaining, Merkle batching, and on-chain anchoring. On-chain bond mechanisms that give third parties economic incentives to audit agent behavior proactively are the most promising next step.

\section*{Acknowledgment}
The authors used AI tools for editorial assistance (drafting, copy-editing, and trimming for length); all technical content, results, and conclusions are the authors' own.

\balance
{\footnotesize
\bibliographystyle{IEEEtran}
\bibliography{references}

% Generated by IEEEtran.bst, version: 1.14 (2015/08/26)
\begin{thebibliography}{10}
\providecommand{\url}[1]{#1}
\csname url@samestyle\endcsname
\providecommand{\newblock}{\relax}
\providecommand{\bibinfo}[2]{#2}
\providecommand{\BIBentrySTDinterwordspacing}{\spaceskip=0pt\relax}
\providecommand{\BIBentryALTinterwordstretchfactor}{4}
\providecommand{\BIBentryALTinterwordspacing}{\spaceskip=\fontdimen2\font plus
\BIBentryALTinterwordstretchfactor\fontdimen3\font minus
  \fontdimen4\font\relax}
\providecommand{\BIBforeignlanguage}[2]{{%
\expandafter\ifx\csname l@#1\endcsname\relax
\typeout{** WARNING: IEEEtran.bst: No hyphenation pattern has been}%
\typeout{** loaded for the language `#1'. Using the pattern for}%
\typeout{** the default language instead.}%
\else
\language=\csname l@#1\endcsname
\fi
#2}}
\providecommand{\BIBdecl}{\relax}
\BIBdecl

\bibitem{microsoft_governance}
\BIBentryALTinterwordspacing
U.~Nagdev and A.~Singh, ``Architecting trust: A {NIST}-based security
  governance framework for {AI} agents,'' Microsoft Tech Community, 2026.
  [Online]. Available:
  \url{https://techcommunity.microsoft.com/blog/microsoftdefendercloudblog/architecting-trust-a-nist-based-security-governance-framework-for-ai-agents/4490556}
\BIBentrySTDinterwordspacing

\bibitem{replit_incident}
\BIBentryALTinterwordspacing
J.~Lemkin, ``{Replit} agent deletes production database during code freeze,'' X
  (formerly Twitter), 2025, corroborated by Fortune (2025-07-23), The Register
  (2025-07-21), and eWeek (2025-07). See also
  \url{https://incidentdatabase.ai/cite/1152/}. [Online]. Available:
  \url{https://x.com/jasonlk/status/1946069562723897802}
\BIBentrySTDinterwordspacing

\bibitem{ncsc_prompt_injection}
\BIBentryALTinterwordspacing
{UK National Cyber Security Centre}, ``Prompt injection is not {SQL} injection
  (it may be worse),'' NCSC Blog, Dec. 2025. [Online]. Available:
  \url{https://www.ncsc.gov.uk/blog-post/prompt-injection-is-not-sql-injection}
\BIBentrySTDinterwordspacing

\bibitem{techradar_second_order}
\BIBentryALTinterwordspacing
S.~Fadilpa\v{s}i\'{c}, ``Second-order prompt injection can turn {AI} into a
  malicious insider,'' TechRadar, Nov. 2025, news article; cited for problem
  context. [Online]. Available:
  \url{https://www.techradar.com/pro/security/second-order-prompt-injection-can-turn-ai-into-a-malicious-insider}
\BIBentrySTDinterwordspacing

\bibitem{swe_agent}
J.~Yang, C.~E. Jimenez, A.~Wettig, K.~Lieret, S.~Yao, K.~Narasimhan, and
  O.~Press, ``{SWE}-agent: Agent-computer interfaces enable automated software
  engineering,'' in \emph{Proc.\ NeurIPS}, 2024.

\bibitem{schneier_kelsey}
\BIBentryALTinterwordspacing
B.~Schneier and J.~Kelsey, ``Secure audit logs to support computer forensics,''
  \emph{ACM Transactions on Information and System Security}, vol.~2, no.~2,
  pp. 159--176, 1999. [Online]. Available:
  \url{https://dl.acm.org/doi/10.1145/317087.317089}
\BIBentrySTDinterwordspacing

\bibitem{cloudtrail_integrity}
\BIBentryALTinterwordspacing
{Amazon Web Services}, ``Validating {CloudTrail} log file integrity,'' AWS
  Documentation, 2024. [Online]. Available:
  \url{https://docs.aws.amazon.com/awscloudtrail/latest/userguide/cloudtrail-log-file-validation-intro.html}
\BIBentrySTDinterwordspacing

\bibitem{rfc6962}
\BIBentryALTinterwordspacing
B.~Laurie, A.~Langley, and E.~Kasper, ``Certificate transparency,'' IETF, RFC
  6962, 2013. [Online]. Available: \url{https://www.rfc-editor.org/rfc/rfc6962}
\BIBentrySTDinterwordspacing

\bibitem{putz_blockchain_logging}
\BIBentryALTinterwordspacing
B.~Putz, F.~Menges, and G.~Pernul, ``A secure and auditable logging
  infrastructure based on a permissioned blockchain,'' \emph{Computers \&
  Security}, vol.~87, p. 101602, 2019. [Online]. Available:
  \url{https://www.sciencedirect.com/science/article/abs/pii/S0167404818313907}
\BIBentrySTDinterwordspacing

\bibitem{crosby_wallach}
S.~A. Crosby and D.~S. Wallach, ``Efficient data structures for tamper-evident
  logging,'' in \emph{Proc.\ USENIX Security Symposium}.\hskip 1em plus 0.5em
  minus 0.4em\relax USENIX, 2009, pp. 317--334.

\bibitem{ietf_agent_audit}
\BIBentryALTinterwordspacing
R.~Sharif, ``Agent audit trail: A standard logging format for autonomous {AI}
  systems,'' IETF, Internet-Draft draft-sharif-agent-audit-trail-00, 2026.
  [Online]. Available:
  \url{https://datatracker.ietf.org/doc/draft-sharif-agent-audit-trail/}
\BIBentrySTDinterwordspacing

\bibitem{auditable_agentic}
C.~C. Phiri, ``Creating characteristically auditable agentic {AI} systems,'' in
  \emph{Proc.\ Intelligent Robotics FAIR (IntRob)}.\hskip 1em plus 0.5em minus
  0.4em\relax ACM, 2025.

\bibitem{rfc8785}
\BIBentryALTinterwordspacing
A.~Rundgren, B.~Jordan, and S.~Erdtman, ``{JSON} canonicalization scheme
  ({JCS}),'' IETF, RFC 8785, 2020. [Online]. Available:
  \url{https://www.rfc-editor.org/rfc/rfc8785}
\BIBentrySTDinterwordspacing

\bibitem{optimism_specs}
\BIBentryALTinterwordspacing
{OP Labs}, ``Optimism fault proof specifications,'' GitHub, 2024. [Online].
  Available:
  \url{https://github.com/ethereum-optimism/specs/blob/main/specs/fault-proof/index.md}
\BIBentrySTDinterwordspacing

\bibitem{rfc3161}
\BIBentryALTinterwordspacing
C.~Adams, P.~Cain, D.~Pinkas, and R.~Zuccherato, ``Internet {X.509} public key
  infrastructure time-stamp protocol ({TSP}),'' IETF, RFC 3161, 2001. [Online].
  Available: \url{https://www.rfc-editor.org/rfc/rfc3161}
\BIBentrySTDinterwordspacing

\bibitem{bellare_yee_forward_integrity}
M.~Bellare and B.~Yee, ``Forward integrity for secure audit logs,'' Department
  of Computer Science and Engineering, University of California at San Diego,
  Tech. Rep. CS98-580, Nov. 1997.

\bibitem{zhao_nitro}
R.~Zhao, M.~Shoaib, V.~T. Hoang, and W.~U. Hassan, ``Rethinking tamper-evident
  logging: A high-performance, co-designed auditing system,'' in \emph{Proc.\
  ACM SIGSAC Conference on Computer and Communications Security (CCS)}.\hskip
  1em plus 0.5em minus 0.4em\relax ACM, 2025.

\bibitem{newman_sigstore}
Z.~Newman, J.~S. Meyers, and S.~Torres-Arias, ``Sigstore: Software signing for
  everybody,'' in \emph{Proc.\ ACM SIGSAC Conference on Computer and
  Communications Security (CCS)}.\hskip 1em plus 0.5em minus 0.4em\relax ACM,
  2022, pp. 2353--2367.

\bibitem{trillian}
\BIBentryALTinterwordspacing
{Google}, ``Trillian: General transparency,'' GitHub, 2024, open-source
  verifiable log infrastructure. [Online]. Available:
  \url{https://github.com/google/trillian}
\BIBentrySTDinterwordspacing

\bibitem{buldas_ksi}
A.~Buldas, A.~Kroonmaa, and R.~Laanoja, ``Keyless signatures' infrastructure:
  How to build global distributed hash-trees,'' in \emph{Secure IT Systems:
  18th Nordic Conference (NordSec)}, ser. LNCS, vol. 8208.\hskip 1em plus 0.5em
  minus 0.4em\relax Springer, 2013, pp. 313--320.

\bibitem{pourmajidi_logchain}
W.~Pourmajidi and A.~V. Miranskyy, ``Logchain: Blockchain-assisted log
  storage,'' in \emph{Proc.\ IEEE 11th International Conference on Cloud
  Computing (CLOUD)}.\hskip 1em plus 0.5em minus 0.4em\relax IEEE, 2018, pp.
  978--982.

\bibitem{jan_blockchain_agent}
S.~Jan, H.~A. Razzaqi, A.~Akarma, and M.~R. Belgaum, ``A blockchain-monitored
  agentic {AI} architecture for trusted perception-reasoning-action
  pipelines,'' in \emph{Proc.\ IEEE Int.\ Conf.\ Computing and Applications
  (ICCA)}.\hskip 1em plus 0.5em minus 0.4em\relax IEEE, 2025, pp. 1--7.

\bibitem{ai_action_ledger}
\BIBentryALTinterwordspacing
Jreamr, ``Built a tamper-evident audit log for {LangChain} agents (early users
  welcome),'' LangChain Forum, 2026, community project; cited for problem
  framing, not as archival source. [Online]. Available:
  \url{https://forum.langchain.com/t/built-a-tamper-evident-audit-log-for-langchain-agents-early-users-welcome/2788}
\BIBentrySTDinterwordspacing

\bibitem{air_project}
\BIBentryALTinterwordspacing
J.~Shotwell, ``{AIR} blackbox: Open-source {AI} governance control plane with
  tamper-evident audit trails,'' GitHub, 2026, open-source project; cited for
  problem framing. See also
  \url{https://news.ycombinator.com/item?id=47061879}. [Online]. Available:
  \url{https://github.com/airblackbox/gateway}
\BIBentrySTDinterwordspacing

\bibitem{prooftrail}
\BIBentryALTinterwordspacing
{ProofTrail}, ``{ProofTrail}: Tamper-proof receipts for {AI} agents,'' Web,
  2026, python library (\texttt{ads-foundation} on PyPI); cited for problem
  framing. [Online]. Available: \url{https://prooftrail.dev/}
\BIBentrySTDinterwordspacing

\bibitem{ojewale_audit_trails}
V.~Ojewale, H.~Suresh, and S.~Venkatasubramanian, ``Audit trails for
  accountability in large language models,'' \emph{arXiv preprint
  arXiv:2601.20727}, 2026.

\bibitem{bcca2025_sok}
\BIBentryALTinterwordspacing
N.~Romandini, C.~Mazzocca, K.~Otsuki, and R.~Montanari, ``{SoK}: Security and
  privacy of {AI} agents for blockchain,'' in \emph{Proc.\ 7th International
  Conference on Blockchain Computing and Applications (BCCA)}.\hskip 1em plus
  0.5em minus 0.4em\relax IEEE, 2025, pp. 708--720. [Online]. Available:
  \url{https://arxiv.org/abs/2509.07131}
\BIBentrySTDinterwordspacing

\bibitem{microsoft_purview}
\BIBentryALTinterwordspacing
{Microsoft}, ``Audit logs for {Copilot} and {AI} applications,'' Microsoft
  Learn, 2026, last updated 2026-02-18. [Online]. Available:
  \url{https://learn.microsoft.com/en-us/purview/audit-copilot}
\BIBentrySTDinterwordspacing

\bibitem{eu_ai_act}
\BIBentryALTinterwordspacing
``Regulation ({EU}) 2024/1689 of the {European Parliament} and of the {Council}
  laying down harmonised rules on artificial intelligence ({AI Act}),''
  Official Journal of the European Union, 2024. [Online]. Available:
  \url{https://eur-lex.europa.eu/legal-content/EN/TXT/HTML/?uri=OJ:L_202401689}
\BIBentrySTDinterwordspacing

\end{thebibliography}
}

\end{document}